\documentclass[twocolumn]{revtex4}

\usepackage{graphicx}
\usepackage{dcolumn}
\usepackage{amsmath}

\begin{document}

\pagestyle{myheadings}
\markright{
Journal of the Physical Society of Japan 81 (2012) 094713}

\title{
Kondo Effect in the Presence of Spin-Orbit Coupling
}

\author{Takashi Yanagisawa
}

\affiliation{Electronics and Photonics Research 
Institute, National Institute of Advanced Industrial Science and Technology (AIST) 
Tsukuba Central 2, 1-1-1 Umezono, Tsukuba 305-8568, Japan
}

\date{Received April 8, 2012; published online September 4, 2012}


\begin{abstract}
Recently, a series of noncentrosymmetric superconductors has been a subject of 
considerable interest since the discovery of superconductivity in CePt$_3$Si.
In noncentrosymmetric materials, the degeneracy of bands is lifted in the presence
of spin-orbit coupling.  This will bring about new effects in the Kondo effect
since the band degeneracy plays an important role in the scattering of electrons
by localized spins.
We investigate the single-impurity Kondo problem in the presence of spin-orbit 
coupling.  We examine the effect of spin-orbit coupling on the scattering of
conduction electrons, by using the Green's function method, for the s-d 
Hamiltonian, with employing a decoupling procedure.  As a result, we obtain a
closed system of equations of Green's functions, from which we can calculate physical
quantities.
The Kondo temperature $T_K$ is estimated from a singularity of Green's functions.
We show that $T_K$ is reduced as the spin-orbit coupling constant $\alpha$ is
increased.
When $2\alpha k_F$ is comparable to or greater than $k_BT_K(\alpha=0)$,
$T_K$ shows an abrupt decrease as a result of the band splitting. 
This suggests a Kondo collapse accompanied with a sharp decrease of $T_K$.
The $\log T$-dependence of the resistivity will be concealed by the spin-orbit
interaction.
\end{abstract}

\maketitle

\section{Introduction}

The Kondo effect has attracted many researchers since the discovery of the solution
of the resistance minimum\cite{kon64,kon69}.
The effect arises from the interactions between a single magnetic atom and
the many electrons in a metal.
Metals, when magnetic atoms are added, and rare earth compounds exhibit many
interesting phenomena that are related to the Kondo effect.
The spin-flip scattering of a conduction-electron spin by a localized impurity
spin gives rise to a term proportional to $\ln T$ in the resistivity.

Superconductors without inversion symmetry have attracted much attention since
the discovery of superconductivity in CePt$_3$Si\cite{bau04}. 
A group of noncentrosymmetric rare-earth compounds has been reported to exhibit 
superconductivity:
for example, Li$_2$Pt$_3$B\cite{yua06,nis07}, CeIrSi$_3$, CeCoGe$_3$, 
CeIrGe$_3$\cite{kim07,mea09,hon10}, and LaNiC$_2$\cite{hil09}.
The absence of spatial inversion yields the splitting of bands due to a spin-orbit
interaction\cite{sam04,has09}.

The influence of the spin-orbit interaction was discussed very recently in
two-dimensional systems starting from the single-impurity Anderson 
model\cite{zar12,zit11,fen11}.
In the conventional Kondo problem, the conduction-electron states with spin up
and down are degenerate.
We expect that the band splitting has a large effect on the Kondo effect,
and is closely related to a multi-channel Kondo problem.
The purpose of this paper is to investigate this subject on the basis of
the s-d Hamiltonian with the spin-orbit interaction of Rashba type in
three dimensions at finite temperature.  
We calculate Green's functions and evaluate the
Kondo temperature $T_K$ from a singularity of them.
We show that $T_K$ is reduced as a result of the band splitting and 
shows a abrupt decrease when $\alpha k_F$ is comparable to $k_BT_K$.

The paper is organized as follows.
In Section II we show the Hamiltonian, and in Section III we derive 
equations for Green's functions.  We obtain an approximate solution
in Section IV.  The Kondo temperature and correction to resistivity
are discussed in subsequent Sections V and VI.
In Section VII we examine the strong limit of the spin-orbit interaction
where the details of calcuations are shown in Appexdix.

\section{Model Hamiltonian}

The Hamiltonian is $H=H_0+H_{sd}=H_K+H_{so}+H_{sd}$ where
\begin{eqnarray}
H_K &=& \sum_{{\bf k}}\xi_{{\bf k}}(c_{{\bf k}\uparrow}^{\dag}c_{{\bf k}\uparrow}
+c_{{\bf k}\downarrow}^{\dag}c_{{\bf k}\downarrow}),\\
H_{so}&=& \sum_{{\bf k}}[ \alpha(ik_x+k_y)c_{{\bf k}\uparrow}^{\dag}
c_{{\bf k}\downarrow}
+\alpha(-ik_x+k_y)c_{{\bf k}\downarrow}^{\dag}c_{{\bf k}\uparrow}],\nonumber\\
&&\\
H_{sd}&=& -\frac{J}{2}\frac{1}{N}\sum_{{\bf k}{\bf k}'}[
S_z(c_{{\bf k}\uparrow}^{\dag}c_{{\bf k}'\uparrow}-
c_{{\bf k}\downarrow}^{\dag}c_{{\bf k}'\downarrow})
+S_{+}c_{{\bf k}\downarrow}^{\dag}c_{{\bf k}'\uparrow}\nonumber\\
&&+S_{-}c_{{\bf k}\uparrow}^{\dag}c_{{\bf k}'\downarrow}].
\end{eqnarray}
$\xi_{{\bf k}}$ is defined by $\xi_{{\bf k}}=\epsilon_{{\bf k}}-\mu$ where 
$\epsilon_{{\bf k}}$ is the
dispersion relation of the conduction electrons and $\mu$ is the chemical
potential.
$c_{{{\bf k}}\sigma}$ and $c_{{{\bf k}}\sigma}^{\dag}$ are annihilation and 
creation operators,
respectively.
We set $H_0=H_K+H_{so}$.
$S_{+}$, $S_{-}$ and $S_z$ denote the operators of the localized spin.
We consider the spin-orbit interaction of Rashba type in $H_{so}$.
$\alpha$ indicates the coupling constant of the spin-orbit interaction.
The term $H_{sd}$ indicates the s-d interaction between the conduction electrons
and the localized spin, with the coupling constant $J$.
$J$ is negative for the antiferromagnetic interaction.

\section{Green's Functions}

First, we define Green's functions of the conduction electrons
\begin{eqnarray}
G_{{\bf k}{\bf k}'\sigma}(\tau)&=&-\langle T_{\tau}c_{{\bf k}\sigma}(\tau)
c_{{\bf k}'\sigma}^{\dag}(0)\rangle,\\
F_{{{\bf k}}{{\bf k}}'}(\tau)&=&-\langle T_{\tau}c_{{{\bf k}}\downarrow}(\tau)
c_{{{\bf k}}'\uparrow}^{\dag}(0)\rangle,
\end{eqnarray}
where $T_{\tau}$ is the time ordering operator.
We note that the spin operators satisfy the following relations:
\begin{eqnarray}
S_{\pm}S_z&=&\mp\frac{1}{2}S_z,~~~S_zS_{\pm}=\pm\frac{1}{2}S_{\pm},\\
S_+S_-&=&\frac{3}{4}+S_z-S_z^2,\\
S_-S_+&=&\frac{3}{4}-S_z-S_z^2.
\end{eqnarray}

We also define Green's functions which include the localized spins as well as the
conduction electron operators.   They are for example, following the notation of 
Zubarev\cite{zub74},
\begin{eqnarray}
\langle\langle S_zc_{{{\bf k}}\uparrow};c_{{{\bf k}}'\uparrow}^{\dag}
\rangle\rangle_{\tau}
&=& -\langle T_{\tau}S_z c_{{{\bf k}}\uparrow}(\tau)c_{{{\bf k}}'\uparrow}^{\dag}(0)\rangle,\\
\langle\langle S_{-}c_{{{\bf k}}\downarrow};c_{{{\bf k}}'\uparrow}^{\dag}
\rangle\rangle_{\tau}
&=& -\langle T_{\tau}S_{-}c_{{{\bf k}}\downarrow}(\tau)c_{{{\bf k}}'\uparrow}^{\dag}(0)
\rangle,\\
\langle\langle S_zc_{{{\bf k}}\downarrow};c_{{{\bf k}}'\uparrow}^{\dag}\rangle\rangle_{\tau}
&=& -\langle T_{\tau}S_z c_{{{\bf k}}\downarrow}(\tau)c_{{{\bf k}}'\uparrow}^{\dag}(0)
\rangle,\\
\langle\langle S_-c_{{{\bf k}}\uparrow};c_{{{\bf k}}'\uparrow}^{\dag}
\rangle\rangle_{\tau}
&=& -\langle T_{\tau}S_- c_{{{\bf k}}\uparrow}(\tau)c_{{{\bf k}}'\uparrow}^{\dag}(0)
\rangle.
\end{eqnarray} 
The Fourier transforms are defined as usual:
\begin{eqnarray}
G_{{{\bf k}}{{\bf k}}'\sigma}(\tau)&=& \frac{1}{\beta}\sum_n e^{-i\omega_n\tau}
G_{{{\bf k}}{{\bf k}}'\sigma}(i\omega_n),\\
F_{{{\bf k}}{{\bf k}}'}(\tau)&=& \frac{1}{\beta}\sum_n e^{-i\omega_n\tau}
F_{{{\bf k}}{{\bf k}}'}(i\omega_n),\\
\langle\langle S_zc_{{{\bf k}}\uparrow};c_{{{\bf k}}'\uparrow}^{\dag}
\rangle\rangle_{\tau}&=&
\frac{1}{\beta}\sum_n e^{-i\omega_n\tau}
\langle\langle S_zc_{{{\bf k}}\uparrow};c_{{{\bf k}}'\uparrow}^{\dag}
\rangle\rangle_{i\omega_n},\\
&& \cdots\cdots .
\end{eqnarray}

From the commutation relations
\begin{eqnarray}
\left[H_0,c_{{{\bf k}}\uparrow}\right]&=& -\xi_{{\bf k}} c_{{{\bf k}}\uparrow}
-\alpha(ik_x+k_y)
c_{{{\bf k}}\downarrow},\\
\left[H_{sd},c_{{{\bf k}}\uparrow}\right]&=& -\frac{J}{2N}\sum_{{{\bf k}}'}
(-S_z c_{{{\bf k}}'\uparrow}
-S_{-}c_{{{\bf k}}'\downarrow}),
\end{eqnarray}
the equation of motion for $G_{{{\bf k}}{{\bf k}}'\uparrow}(\tau)$ reads
\begin{eqnarray}
\frac{\partial}{\partial\tau}G_{{{\bf k}}{{\bf k}}'\uparrow}(\tau)&=& 
-\delta(\tau)\delta_{{{\bf k}}{{\bf k}}'}
-\xi_{{\bf k}} G_{{{\bf k}}{{\bf k}}'\uparrow}(\tau)\nonumber\\
&&- \alpha(ik_x+k_y)F_{{{\bf k}}{{\bf k}}'}(\tau)\nonumber\\
&+&\frac{J}{2N}\sum_{{{\bf q}}}[ 
\langle\langle S_zc_{{{\bf q}}\uparrow};c_{{{\bf k}}'\uparrow}^{\dag}
\rangle\rangle_{\tau} 
+\langle\langle S_-c_{q\downarrow};c_{{{\bf k}}'\uparrow}^{\dag}\rangle\rangle_{\tau} ].
\nonumber\\
\end{eqnarray}
Similarly, the equation of motion for $F_{{{\bf k}}{{\bf k}}'}$ is
\begin{eqnarray}
\frac{\partial}{\partial\tau}F_{{{\bf k}}{{\bf k}}'}(\tau)&=& -\xi_{{\bf k}} 
F_{{{\bf k}}{{\bf k}}'}
-\alpha(-ik_x+k_y)G_{{{\bf k}}{{\bf k}}'\uparrow}(\tau)\nonumber\\
&-& \frac{J}{2N}\sum_q [
\langle\langle S_zc_{{{\bf q}}\downarrow};c_{{{\bf k}}'\uparrow}^{\dag}
\rangle\rangle_{\tau} 
- \langle\langle S_+c_{{{\bf q}}\uparrow};c_{{{\bf k}}'\uparrow}^{\dag}
\rangle\rangle_{\tau} ].\nonumber\\
\end{eqnarray}
We define
\begin{eqnarray}
\Gamma_{{{\bf k}}{{\bf k}}'}(\tau)&=&\frac{1}{\beta}\sum_n e^{-i\omega_n}
\Gamma_{{{\bf k}}{{\bf k}}'}(i\omega_n)\nonumber\\
&=& \langle\langle S_zc_{{{\bf k}}\uparrow};c_{{{\bf k}}'\uparrow}^{\dag}
\rangle\rangle_{\tau} 
+ \langle\langle S_-c_{{{\bf k}}\downarrow};c_{{{\bf k}}'\uparrow}^{\dag}
\rangle\rangle_{\tau},\\
\Phi_{{{\bf q}}{{\bf k}}}(\tau)&=& \frac{1}{\beta}\sum_n e^{-i\omega_n\tau}
\Phi_{{{\bf q}}{{\bf k}}}(i\omega_n)\nonumber\\
&=&\langle\langle S_zc_{{{\bf q}}\downarrow}-S_+c_{{{\bf q}}\uparrow};
c_{{{\bf k}}\uparrow}^{\dag}
\rangle\rangle,
\end{eqnarray}
then we obtain
\begin{eqnarray}
(i\omega_n-\xi_{{\bf k}})G_{{{\bf k}}{{\bf k}}'\uparrow}(i\omega_n)&=& 
\delta_{{{\bf k}}{{\bf k}}'}
+\alpha(ik_x+k_y)F_{{{\bf k}}{{\bf k}}'}(i\omega_n)\nonumber\\
&-& \frac{J}{2N}\sum_{{{\bf q}}{{\bf k}}'}\Gamma_{{{\bf q}}{{\bf k}}'}(i\omega_n),\\
(i\omega_n-\xi_{{\bf k}})F_{{{\bf k}}{{\bf k}}'}(i\omega_n)&=& 
\alpha(-ik_x+k_y)G_{{{\bf k}}{{\bf k}}'\uparrow}(i\omega_n)\nonumber\\
&+&\frac{J}{2N}\sum_q \Phi_{{{\bf q}}{{\bf k}}'}(i\omega_n).
\end{eqnarray}
To obtain the solution to the equations above, we need Green's functions in 
eqs.(9)-(12).
The equations of motion for 
$\langle\langle S_zc_{{{\bf k}}\uparrow};c_{{{\bf k}}'\uparrow}^{\dag}\rangle\rangle$
and $\langle\langle S_-c_{{{\bf k}}\downarrow};c_{{{\bf k}}'\uparrow}^{\dag}
\rangle\rangle$ are
\begin{eqnarray}
&&(i\omega-\xi_{{\bf k}})
\langle\langle S_zc_{{{\bf k}}\uparrow};c_{{{\bf k}}'\uparrow}^{\dag}
\rangle\rangle_{i\omega}\nonumber\\
&=& \langle S_z\rangle\delta_{{{\bf k}}{{\bf k}}'}
+ \alpha(ik_x+k_y)
\langle\langle S_zc_{{{\bf k}}\downarrow};c_{{{\bf k}}'\uparrow}^{\dag}
\rangle\rangle_{i\omega}
\nonumber\\
&-& \frac{J}{2N}\sum_{{\bf q}} \Big[
\langle\langle S_z^2c_{{{\bf q}}\uparrow};c_{{{\bf k}}'\uparrow}^{\dag}
\rangle\rangle_{i\omega}
+\frac{1}{2}
\langle\langle S_-c_{{{\bf q}}\downarrow};c_{{{\bf k}}'\uparrow}^{\dag}
\rangle\rangle_{i\omega}\Big]
\nonumber\\
&-& \frac{J}{2N}\sum_{{{\bf q}}{{\bf q}}'}\Big[
\langle\langle S_+c_{{{\bf k}}\uparrow}c_{{{\bf q}}\downarrow}^{\dag}
c_{{{\bf q}}'\uparrow};
c_{{{\bf k}}'\uparrow}^{\dag}\rangle\rangle_{i\omega}\nonumber\\
&-& \langle\langle S_-c_{{{\bf k}}\uparrow}c_{{{\bf q}}\uparrow}^{\dag}
c_{{{\bf q}}'\downarrow};
c_{{{\bf k}}'\uparrow}^{\dag}\rangle\rangle_{i\omega} \Big],\nonumber\\
\end{eqnarray}
\begin{eqnarray}
&&(i\omega-\xi_{{\bf k}})\langle\langle S_-c_{{{\bf k}}\downarrow};
c_{{{\bf k}}'\uparrow}^{\dag}
\rangle\rangle_{i\omega}\nonumber\\
&=& \alpha(-ik_x+k_y)\langle\langle S_-c_{{{\bf k}}\uparrow};
c_{{{\bf k}}'\uparrow}^{\dag}
\rangle\rangle_{i\omega}\nonumber\\
&-&\frac{J}{4N}\sum_{{{\bf q}}{{\bf q}}'}\langle\langle S_-c_{{{\bf q}}'\downarrow};
c_{{{\bf k}}'\uparrow}^{\dag}
\rangle\rangle_{i\omega}\nonumber\\
&+&\frac{J}{2N}\sum_{{{\bf q}}{{\bf q}}'}\Big[ \langle\langle S_-c_{{{\bf k}}\downarrow}
c^{\dag}_{{{\bf q}}\uparrow}
c_{{{\bf q}}'\uparrow};c_{{{\bf k}}'\uparrow}^{\dag}\rangle\rangle_{i\omega}\nonumber\\
&&- \langle\langle S_-c_{{{\bf k}}\downarrow}c^{\dag}_{{{\bf q}}\downarrow}
c_{{{\bf q}}'\downarrow};c_{{{\bf k}}'\uparrow}^{\dag}\rangle\rangle_{i\omega}
-2\langle\langle S_zc_{{{\bf k}}\downarrow}c^{\dag}_{{{\bf q}}\downarrow}
c_{{{\bf q}}'\uparrow};c_{{{\bf k}}'\uparrow}^{\dag}\rangle\rangle_{i\omega}\Big]
\nonumber\\
&&-\frac{J}{2N}\sum_{{{\bf q}}'}\langle\langle S_+S_-c_{{{\bf q}}'\uparrow};
c_{{{\bf k}}'\uparrow}^{\dag}
\rangle\rangle_{i\omega}.
\end{eqnarray}
We use the commutation relation $S_+S_-=3/4+S_z-S_z^2$ to obtain
\begin{eqnarray}
&&(i\omega-\xi_{{\bf k}})\Gamma_{{{\bf k}}{{\bf k}}'}(i\omega)\nonumber\\
&=& \delta_{{{\bf k}}{{\bf k}}'}\langle S_z\rangle
+\alpha(ik_x+k_y)
\langle\langle S_zc_{{{\bf k}}\downarrow};c_{{{\bf k}}'\uparrow}^{\dag}
\rangle\rangle_{i\omega}\nonumber\\
&+& \alpha(-ik_x+k_y)\langle\langle
S_-c_{{{\bf k}}\uparrow};c_{{{\bf k}}'\uparrow}^{\dag}\rangle\rangle_{i\omega}
\nonumber\\
&-&\frac{J}{2N}\sum_{{\bf q}}\Big[ \frac{3}{4}\langle\langle c_{{{\bf q}}\uparrow};
c_{{{\bf k}}'\uparrow}^{\dag}\rangle\rangle_{i\omega} 
+\Gamma_{{{\bf q}}{{\bf k}}'}(i\omega)\Big]\nonumber\\
&-&\frac{J}{2N}\sum_{{{\bf q}}{{\bf q}}'}\Big[
\langle\langle S_+c_{{{\bf k}}\uparrow}c_{{{\bf q}}\downarrow}^{\dag}
c_{{{\bf q}}'\uparrow};
c_{{{\bf k}}'\uparrow}^{\dag}\rangle\rangle_{i\omega} \nonumber\\
&-& \langle\langle S_-c_{{{\bf k}}\uparrow}c_{{{\bf q}}\uparrow}^{\dag}
c_{{{\bf q}}'\downarrow};
c_{{{\bf k}}'\uparrow}^{\dag}\rangle\rangle_{i\omega}
- \langle\langle S_-c_{{{\bf k}}\downarrow}c_{{{\bf q}}\uparrow}^{\dag}
c_{{{\bf q}}'\uparrow};
c_{{{\bf k}}'\uparrow}^{\dag}\rangle\rangle_{i\omega}\nonumber\\
&+& \langle\langle S_-c_{{{\bf k}}\downarrow}c_{{{\bf q}}\downarrow}^{\dag}
c_{{{\bf q}}'\downarrow};
c_{{{\bf k}}'\uparrow}^{\dag}\rangle\rangle_{i\omega}\nonumber\\
&+& 2\langle\langle S_zc_{{{\bf k}}\downarrow}c_{{{\bf q}}\downarrow}^{\dag}
c_{{{\bf q}}'\uparrow};
c_{{{\bf k}}'\uparrow}^{\dag}\rangle\rangle_{i\omega}\Big].
\end{eqnarray}
Here we assume that $\langle S_z\rangle=0$.
Now we need $\langle\langle S_zc_{{{\bf k}}\downarrow};
c_{{{\bf k}}'\uparrow}^{\dag}\rangle\rangle$
and $\langle\langle S_-c_{{{\bf k}}\uparrow};c_{{{\bf k}}'\uparrow}^{\dag}
\rangle\rangle$ to
obtain a solution for $\Gamma_{{{\bf k}}{{\bf k}}'}$.
The equations for 
$\langle\langle S_zc_{{{\bf k}}\downarrow};c_{{{\bf k}}'\uparrow}^{\dag}\rangle\rangle$
and $\langle\langle S_-c_{{{\bf k}}\uparrow};c_{{{\bf k}}'\uparrow}^{\dag}\rangle\rangle$ read
\begin{eqnarray}
&&(i\omega-\xi_{{\bf k}})
\langle\langle S_zc_{{{\bf k}}\downarrow};c_{{{\bf k}}'\uparrow}^{\dag}
\rangle\rangle_{i\omega}\nonumber\\
&=& \alpha (-ik_x+k_y)
\langle\langle S_zc_{{{\bf k}}\downarrow};c_{{{\bf k}}'\uparrow}^{\dag}
\rangle\rangle_{i\omega}\nonumber\\
&+& \frac{J}{2N}\sum_q
\Big[\langle\langle S_z^2c_{{{\bf q}}\downarrow};c_{{{\bf k}}'\uparrow}^{\dag}
\rangle\rangle_{i\omega}
-\frac{1}{2}
\langle\langle S_+c_{{{\bf q}}\uparrow};c_{{{\bf k}}'\uparrow}^{\dag}
\rangle\rangle_{i\omega}\Big]\nonumber\\
&-&\frac{J}{2N}\sum_{{{\bf q}}{{\bf q}}'}\Big[
\langle\langle S_+c_{{{\bf q}}\downarrow}^{\dag}c_{{{\bf q}}'\downarrow}
c_{{{\bf k}}\downarrow};
c_{{{\bf k}}'\uparrow}^{\dag}\rangle\rangle_{i\omega}
\nonumber\\
&-& \langle\langle S_-c_{{{\bf q}}\uparrow}^{\dag}c_{{{\bf q}}'\downarrow}
c_{{{\bf k}}\downarrow};
c_{{{\bf k}}'\uparrow}^{\dag}\rangle\rangle_{i\omega}\Big],
\end{eqnarray}

\begin{eqnarray}
&&(i\omega-\xi_{{\bf k}})
\langle\langle S_-c_{{{\bf k}}\uparrow};c_{{{\bf k}}'\uparrow}^{\dag}
\rangle\rangle_{i\omega}\nonumber\\
&=& -\delta_{{{\bf k}}{{\bf k}}'}\langle S_-\rangle
+\alpha (ik_x+k_y)\langle\langle S_-c_{{{\bf k}}\downarrow};c_{{{\bf k}}'\uparrow}^{\dag}
\rangle\rangle_{i\omega}\nonumber\\
&+&\frac{J}{4N}\sum_{{{\bf q}}'}
\langle\langle S_-c_{{{\bf q}}'\uparrow};c_{{{\bf k}}'\uparrow}^{\dag}
\rangle\rangle_{i\omega}
\nonumber\\
&+&\frac{J}{2N}\sum_{{{\bf q}}{{\bf q}}'}\Big[
\langle\langle S_-c_{{{\bf k}}\uparrow}c_{{{\bf q}}\downarrow}^{\dag}
c_{{{\bf q}}'\downarrow};
c_{{{\bf k}}'\uparrow}^{\dag}\rangle\rangle_{i\omega}\nonumber\\
&-&\langle\langle S_-c_{{{\bf k}}\uparrow}c_{{{\bf q}}\uparrow}^{\dag}
c_{{{\bf q}}'\uparrow};
c_{{{\bf k}}'\uparrow}^{\dag}\rangle\rangle_{i\omega}
+ 2\langle\langle S_zc_{{{\bf q}}\downarrow}^{\dag}c_{{{\bf q}}'\downarrow}
c_{{{\bf k}}\uparrow};
c_{{{\bf k}}'\uparrow}^{\dag}\rangle\rangle_{i\omega}\Big].\nonumber\\
\end{eqnarray}

\section{Approximate Solution}

We assume that the spin-orbit coupling $\alpha$ is small and we keep terms
up to the order of $\alpha$.  We adopt the approximation that
\begin{eqnarray}
\langle\langle S_zc_{{{\bf k}}\downarrow};c_{{{\bf k}}'\uparrow}^{\dag}
\rangle\rangle_{i\omega}
&=& \frac{\alpha(-ik_x+k_y)}{i\omega-\xi_{{\bf k}}}
\langle\langle S_zc_{{{\bf k}}\uparrow};c_{{{\bf k}}'\uparrow}^{\dag}
\rangle\rangle_{i\omega},\nonumber\\
&& \\
\langle\langle S_-c_{{{\bf k}}\uparrow};c_{{{\bf k}}'\uparrow}^{\dag}
\rangle\rangle_{i\omega}
&=& \frac{\alpha(ik_x+k_y)}{i\omega-\xi_{{\bf k}}}
\langle\langle S_-c_{{{\bf k}}\downarrow};c_{{{\bf k}}'\uparrow}^{\dag}
\rangle\rangle_{i\omega}.
\end{eqnarray}
This means that we have neglected the terms of the order of $J\alpha$ in the
right-hand side.
Then we obtain
\begin{eqnarray}
&&\left(i\omega-\xi_{{\bf k}}-\frac{\alpha^2(k_x^2+k_y^2)}{i\omega-\xi_{{\bf k}}}\right)
\Gamma_{{{\bf k}}{{\bf k}}'}(i\omega)\nonumber\\
&=& -\frac{J}{2N}\sum_{{\bf q}}\Big[ \frac{3}{4}
\langle\langle c_{{{\bf q}}\uparrow};c_{{{\bf k}}'\uparrow}^{\dag}
\rangle\rangle_{i\omega}
+\Gamma_{{{\bf q}}{{\bf k}}'}(i\omega)\Big]\nonumber\\
&-& \frac{J}{2N}\sum_{{{\bf q}}{{\bf q}}'}\Big[ 
\langle\langle S_+c_{{{\bf k}}\uparrow}c_{{{\bf q}}\downarrow}^{\dag}
c_{{{\bf q}}'\uparrow};
c_{{{\bf k}}'\uparrow}^{\dag}\rangle\rangle_{i\omega}\nonumber\\
&-& \langle\langle S_-c_{{{\bf k}}\uparrow}c_{{{\bf q}}\downarrow}^{\dag}
c_{{{\bf q}}'\downarrow};
c_{{{\bf k}}'\uparrow}^{\dag}\rangle\rangle_{i\omega}
-\langle\langle S_-c_{{{\bf k}}\downarrow}c_{{{\bf q}}\uparrow}^{\dag}
c_{{{\bf q}}'\uparrow};
c_{{{\bf k}}'\uparrow}^{\dag}\rangle\rangle_{i\omega}\nonumber\\
&+& \langle\langle S_-c_{{{\bf k}}\downarrow}c_{{{\bf q}}\downarrow}^{\dag}
c_{{{\bf q}}'\downarrow};
c_{{{\bf k}}'\uparrow}^{\dag}\rangle\rangle_{i\omega}
+2\langle\langle S_zc_{{{\bf k}}\downarrow}c_{{{\bf q}}\downarrow}^{\dag}
c_{{{\bf q}}'\downarrow};
c_{{{\bf k}}'\uparrow}^{\dag}\rangle\rangle_{i\omega}\Big]\nonumber\\
\end{eqnarray}
We use the same approximation in the right-hand side of eq.(23), that is,
$(i\omega-\xi_{{\bf k}})F_{{{\bf k}}{{\bf k}}'}(i\omega)=\alpha(-ik_x+k_y)
G_{{{\bf k}}{{\bf k}}'\uparrow}(i\omega)$,
 and we have
\begin{eqnarray}
(i\omega-\xi_{{\bf k}})G_{{{\bf k}}{{\bf k}}'\uparrow}(i\omega)
&=& \delta_{{{\bf k}}{{\bf k}}'}-\frac{J}{2N}\sum_{{{\bf p}}}
\Gamma_{{{\bf p}}{{\bf k}}'}(i\omega)\nonumber\\
&+&\frac{\alpha^2(k_x^2+k_y^2)}{i\omega-\xi_{{\bf k}}}
G_{{{\bf k}}{{\bf k}}'\uparrow}(i\omega)
\end{eqnarray}
Here, we employ the decoupling approximation procedure for Green's
functions\cite{nag65,ham67}.
Many-body Green's functions are approximated as follows.
\begin{eqnarray}
\langle\langle S_-c_{{{\bf k}}\uparrow}c_{{{\bf q}}\uparrow}^{\dag}
c_{{{\bf q}}'\downarrow};
c_{{{\bf k}}'\uparrow}^{\dag}\rangle\rangle_{i\omega}
&\approx& \langle c_{{{\bf k}}\uparrow}c_{{{\bf q}}\uparrow}^{\dag}\rangle
\langle\langle S_-c_{{{\bf q}}'\downarrow};c_{{{\bf k}}'\uparrow}^{\dag}
\rangle\rangle_{i\omega}\nonumber\\
&+& \langle S_-c_{{{\bf q}}\uparrow}^{\dag}c_{{{\bf q}}'\downarrow}\rangle
\langle\langle c_{{{\bf k}}\uparrow};c_{{{\bf k}}'\uparrow}^{\dag}
\rangle\rangle_{i\omega},\nonumber\\
&&\\
\langle\langle S_+c_{{{\bf k}}\uparrow}c_{{{\bf q}}\downarrow}^{\dag}
c_{{{\bf q}}'\uparrow};
c_{{{\bf k}}'\uparrow}^{\dag}\rangle\rangle_{i\omega}
&\approx& \langle S_+c_{{{\bf q}}\downarrow}^{\dag}c_{{{\bf q}}'\uparrow}\rangle
\langle\langle c_{{{\bf k}}\uparrow};c_{{{\bf k}}'\uparrow}^{\dag}
\rangle\rangle_{i\omega}\nonumber\\
&-& \langle S_+c_{{{\bf q}}\downarrow}^{\dag}c_{{{\bf k}}'\uparrow}\rangle
\langle\langle c_{{{\bf q}}'\uparrow};c_{{{\bf k}}'\uparrow}^{\dag}
\rangle\rangle_{i\omega},\nonumber\\
&&\\
\langle\langle S_zc_{{{\bf k}}\downarrow}c_{{{\bf q}}\downarrow}^{\dag}
c_{{{\bf q}}'\uparrow};
c_{{{\bf k}}'\uparrow}^{\dag}\rangle\rangle_{i\omega}
&\approx& \langle c_{{{\bf k}}\downarrow}c_{{{\bf q}}\downarrow}^{\dag}\rangle
\langle\langle S_zc_{{{\bf q}}'\uparrow};c_{{{\bf k}}'\uparrow}^{\dag}
\rangle\rangle_{i\omega}\nonumber\\
&-& \langle S_zc_{{{\bf q}}\downarrow}^{\dag}c_{{{\bf k}}\downarrow}\rangle
\langle\langle c_{{{\bf q}}'\uparrow};c_{{{\bf k}}'\uparrow}^{\dag}
\rangle\rangle_{i\omega},\nonumber\\
&&\\
\langle\langle S_-c_{{{\bf k}}\downarrow}c_{{{\bf q}}\downarrow}^{\dag}
c_{{{\bf q}}'\downarrow};
c_{{{\bf k}}'\uparrow}^{\dag}\rangle\rangle_{i\omega}
&\approx& \langle c_{{{\bf q}}\downarrow}c_{{{\bf q}}'\downarrow}^{\dag}\rangle
\langle\langle S_-c_{{{\bf k}}\downarrow};c_{{{\bf k}}'\uparrow}^{\dag}
\rangle\rangle_{i\omega}\nonumber\\
&+& \langle c_{{{\bf k}}\downarrow}c_{{{\bf q}}\downarrow}^{\dag}\rangle
\langle\langle S_-c_{{{\bf q}}'\downarrow};c_{{{\bf k}}'\uparrow}^{\dag}
\rangle\rangle_{i\omega}.\nonumber\\
\end{eqnarray} 
We define
\begin{eqnarray}
n_{{\bf k}} &=& \sum_{{\bf q}}\langle c_{{{\bf q}}\uparrow}^{\dag}
c_{{{\bf k}}\uparrow}\rangle
= \sum_{{\bf q}}\langle c_{{{\bf q}}\downarrow}^{\dag}c_{{{\bf k}}\downarrow}\rangle,\\
m_k &=& 3\sum_{{\bf q}}\langle S_-c_{{{\bf q}}\uparrow}^{\dag}c_{{{\bf k}}\downarrow}
\rangle
= 2\sum_{{\bf q}}( \langle S_zc_{{{\bf q}}\uparrow}^{\dag}c_{{{\bf k}}\uparrow}\rangle
+\langle S_{-}c_{{{\bf q}}\uparrow}^{\dag}c_{{{\bf k}}\downarrow}\rangle ).\nonumber\\
\end{eqnarray}
We used the relation obtained from the rotational symmetry in the spin space,
\begin{equation}
\langle S_-c_{{{\bf q}}\uparrow}^{\dag}c_{{{\bf q}}'\downarrow}\rangle
= \langle S_+c_{{{\bf q}}\downarrow}^{\dag}c_{{{\bf q}}'\uparrow}\rangle
=2\langle S_zc_{{{\bf q}}\uparrow}^{\dag}c_{{{\bf q}}'\uparrow}\rangle
= -2\langle S_zc_{{{\bf q}}\downarrow}^{\dag}c_{{{\bf q}}'\downarrow}\rangle.
\end{equation}
Then, after the analytic continuation $i\omega\rightarrow \omega+i\delta$,
 we have
\begin{eqnarray}
&&\left( \omega-\xi_{{{\bf k}}}-\frac{\alpha^2k_{\perp}^2}{\omega-\xi_{{{\bf k}}}}\right)
\Gamma_{{{\bf k}}{{\bf k}}'}(\omega) 
+ \left(\frac{3}{4}-m_{{{\bf k}}}\right)\nonumber\\
&\times& \frac{J}{2N}\sum_{{\bf q}} 
G_{{{\bf q}}{{\bf k}}'}(\omega)
+ \left(n_{{{\bf k}}}-\frac{1}{2}\right)\frac{J}{N}\sum_{{\bf q}}
\Gamma_{{{\bf q}}{{\bf k}}'}(\omega)
= 0,\nonumber\\
\end{eqnarray}
\begin{equation}
\left( \omega-\xi_{{{\bf k}}}-\frac{\alpha^2k_{\perp}^2}{\omega-\xi_{{{\bf k}}}}\right)
G_{{{\bf k}}{{\bf k}}'}(\omega)
+ \frac{J}{2N}\sum_{{\bf q}} \Gamma_{{{\bf q}}{{\bf k}}'}(\omega)
= \delta_{{{\bf k}}{{\bf k}}'},
\end{equation}
where we set $k_{\perp}^2= k_x^2+k_y^2$.
Then, we obtain from eqs.(41) and (42) that
\begin{eqnarray}
\Gamma_{{{\bf k}}{{\bf k}}'}(\omega)&=& G_{{\bf k}}^0(\omega)\left(m_{{\bf k}}
-\frac{3}{4}\right)
\frac{J}{2N}G_{{{\bf k}}'}^0(\omega)\nonumber\\
&-& G_{{\bf k}}^0(\omega)\Big[ \left(n_{{\bf k}}-\frac{1}{2}\right)J
+\left(m_{{\bf k}}-\frac{3}{4}\right)
\left(\frac{J}{2}\right)^2F(\omega)  \Big]\nonumber\\
&\times& \frac{1}{N}\sum_{{\bf q}}\Gamma_{{{\bf q}}{{\bf k}}}(\omega),\nonumber\\
\end{eqnarray}
where
\begin{eqnarray}
G_{{\bf k}}^0(\omega)&=&\frac{1}{2}\left( \frac{1}{\omega-\xi_{{\bf k}}+\alpha k_{\perp}}
+\frac{1}{\omega-\xi_{{\bf k}}-\alpha k_{\perp}} \right)\\
F(\omega) &=& \frac{1}{N}\sum_{{\bf k}} G_{{\bf k}}^0(\omega),\\
G(\omega) &=& \frac{1}{N}\sum_{{\bf k}} \left(n_{{\bf k}}-\frac{1}{2}\right)
G_{{\bf k}}^0(\omega),\\
\Gamma(\omega) &=& \frac{1}{N}\sum_{{\bf k}}\left( m_{{\bf k}}-\frac{3}{4}\right)
G_{{\bf k}}^0(\omega). 
\end{eqnarray}
Because of
\begin{equation}
\frac{1}{N}\sum_{{\bf q}}\Gamma_{{{\bf q}}{{\bf k}}}(\omega) 
= \frac{J}{2N}\Gamma(\omega)
G_{{\bf k}}^0(\omega)
\frac{1}{1+JG(\omega)+(J/2)^2\Gamma(\omega)F(\omega)},
\end{equation}
we obtain
\begin{eqnarray}
\Gamma_{{{\bf k}}{{\bf k}}'}(\omega) &=& \frac{J}{2N}G_{{\bf k}}^0(\omega)
G_{{{\bf k}}'}^0(\omega)
\Big[ \left(m_{{\bf k}}-\frac{3}{4}\right)(1+JG(\omega))\nonumber\\
&-&\left(n_{{\bf k}}-\frac{1}{2}\right)J\Gamma(\omega)\Big]\nonumber\\
&\times& \frac{1}{1+JG(\omega)+(J/2)^2\Gamma(\omega)F(\omega)},
\end{eqnarray}
\begin{eqnarray}
G_{{{\bf k}}{{\bf k}}'}(\omega) &=& \delta_{{{\bf k}}{{\bf k}}'}G_{{\bf k}}^0(\omega)
-\frac{J^2}{4N}\Gamma(\omega)
G_{{\bf k}}^0(\omega)G_{{{\bf k}}'}^0(\omega)\nonumber\\
&\times& \frac{1}{1+JG(\omega)+(J/2)^2\Gamma(\omega)F(\omega)}\nonumber\\
&=& \delta_{{{\bf k}}{{\bf k}}'}G_{{\bf k}}^0(\omega)+\frac{J}{N}
G_{{\bf k}}^0(\omega)G_{{{\bf k}}'}^0(\omega)t(\omega),
\end{eqnarray}
where we defined
\begin{equation}
t(\omega)= -\frac{J}{4}\frac{\Gamma(\omega)}{1+JG(\omega)
+(J/2)^2\Gamma(\omega)F(\omega)}.
\end{equation}
$m_{{\bf k}}$ is given by
\begin{eqnarray}
m_{{\bf k}}^* &=& 2\sum_{{\bf q}}( \langle S_zc_{{{\bf k}}\uparrow}^{\dag}
c_{{{\bf q}}\uparrow}\rangle
\langle S_-c_{{{\bf k}}\uparrow}^{\dag}c_{{{\bf q}}\downarrow}\rangle )
= 2\sum_{{\bf q}}\Gamma_{{{\bf q}}{{\bf k}}}(\tau=-0)\nonumber\\
&=& \frac{2}{\beta}\sum_{{{\bf q}}\omega_n}e^{i\omega_n\delta}
\Gamma_{{{\bf q}}{{\bf k}}}(i\omega_n),
\end{eqnarray}
where $\delta$ is an infinitesimal constant.  Because $m_{{\bf k}}$ is real, we
obtain 
\begin{equation}
m_{{\bf k}} = -4\frac{1}{\beta}\sum_{\omega_n}e^{i\omega_n\delta}G_{{\bf k}}^0
(i\omega_n)
t(i\omega_n).
\end{equation}
Similarly we have
\begin{equation}
n_{{\bf k}}= \frac{1}{\beta}\sum_{\omega_n}G_{{\bf k}}^0(i\omega_n)
(1+JF(i\omega_n)t(i\omega_n)).
\end{equation}

\section{Kondo Temperature}

A singularity of $t(\omega)$ determines the characteristic temperature of
the system.
We investigate the high-temperature region where $m_{{\bf k}}=0$.  Then,
\begin{equation}
\Gamma(\omega)= -\frac{3}{4}F(\omega).
\end{equation}
We obtain for ${{\bf k}}={{\bf k}}'$
\begin{eqnarray}
G_{{{\bf k}}{{\bf k}}}(\omega)^{-1} &=& G_{{\bf k}}^0(\omega)^{-1}-\frac{3J^2}{16N}
\frac{F(\omega)}{1+JG(\omega)}+O(J^4). \nonumber\\
\end{eqnarray}
The Kondo temperature $T_K$ is determined by the vanishing of the denominator
in this equation:
\begin{eqnarray}
&&1-J\frac{1}{2N}\sum_{{\bf k}}\left( \frac{1}{\omega-\xi_{{\bf k}}+\alpha k_{\perp}}
+\frac{1}{\omega-\xi_{{\bf k}}-\alpha k_{\perp}}\right)\nonumber\\
&\times& \frac{1}{4}\left( \tanh\left(\frac{\xi_{{\bf k}}-\alpha k_{\perp}}{2T_K}\right)
+\tanh\left(\frac{\xi_{{\bf k}}+\alpha k_{\perp}}{2T_K}\right) \right)=0,
\nonumber\\
\end{eqnarray}
where
\begin{equation}
\xi_{{\bf k}}= \frac{1}{2m}(k_{\perp}^2+k_z^2)-\mu.
\end{equation}
We have used 
$n_{\bf k}= (f(\xi_k-\alpha k_{\perp})+f(\xi_k+\alpha k_{\perp}))/2$ by
neglecting the term of the order of $J^2$.
By using the expansion,
\begin{equation}
\tanh\left(\frac{z}{2}\right)= \sum_{n=-\infty}^{\infty}\frac{1}{z-i\pi(2n+1)},
\end{equation}
the equation for $T_K$ is
\begin{eqnarray}
1&=& J\sum_{n=-\infty}^{\infty}\frac{1}{8(2\pi)^2}\sqrt{\frac{2m}{\mu}}\int_0^K
dk_{\perp}k_{\perp}
\Big[ 2\frac{i\pi T_K{\rm sign}(2n+1)}{\omega-i\pi(2n+1)T_K}\nonumber\\
&+& \frac{i\pi T_K{\rm sign}(2n+1)}{\omega+2\alpha k_{\perp}-i\pi(2n+1)T_K}\nonumber\\
&+&\frac{i\pi T_K {\rm sign}(2n+1)}{\omega-2\alpha k_{\perp}-i\pi(2n+1)T_K} \Big],
\end{eqnarray}
where $K$ is a cutoff and we use an approximate expression
\begin{eqnarray}
&&\int_{-K}^Kdk_z \frac{1}{\omega-k_{\perp}^2/(2m)+\alpha k_{\perp}-k_z^2/(2m)+\mu}
\nonumber\\
 &\times&
\frac{1}{k_{\perp}^2/(2m)+k_z^2/(2m)+\alpha k_{\perp}-\mu-i\pi(2n+1)T_K}\nonumber\\
&\approx& \sqrt{\frac{2m}{\mu}}i\pi{\rm sign}(2n+1)T_K
\frac{1}{\omega+2\alpha k_{\perp}-i\pi(2n+1)T_K}.\nonumber\\
\end{eqnarray}
We set an cutoff $n_0\equiv D/(2\pi T_K)$ in the summation with respect to $n$.
By using the formula for the digamma function,
\begin{equation}
\sum_{n=0}^{n_0}\frac{1}{n+\frac{1}{2}+x} = \psi\left(\frac{1}{2}+x+n_0\right)
-\psi\left(\frac{1}{2}+x\right),
\end{equation}
we obtain
\begin{eqnarray}
1&=&|J|\frac{1}{32\pi^2}\sqrt{\frac{2m}{\mu}}\int_0^Kdk_{\perp}k_{\perp}
\Big[ 4\log\left(\frac{2e^{\gamma}D}{\pi T_K}\right)
+ 2\psi\left(\frac{1}{2}\right)\nonumber\\
&-&\frac{1}{2}\psi\left(\frac{1}{2}-\frac{\omega+2\alpha k_{\perp}}{i2\pi T_K}\right)
\nonumber\\
&-&\frac{1}{2}\psi\left(\frac{1}{2}+\frac{\omega+2\alpha k_{\perp}}{i2\pi T_K}\right)
-\frac{1}{2}\psi\left(\frac{1}{2}-\frac{\omega-2\alpha k_{\perp}}{i2\pi T_K}\right)
\nonumber\\
&-&\frac{1}{2}\psi\left(\frac{1}{2}+\frac{\omega-2\alpha k_{\perp}}{i2\pi T_K}\right)
\Big].
\end{eqnarray}
We set $\mu=k_F^2/(2m)$ and $K=2k_F$, and expand the digamma function in
terms of $\alpha k_{\perp}/(2\pi T_K)$.  For $\omega=0$, we have
\begin{eqnarray}
1&=&|J|\frac{mk_F}{2\pi^2}\Big[ \log\left(\frac{2e^{\gamma}D}{\pi T_K}\right)
-\frac{7\zeta(3)}{2\pi^2}\left(\frac{2\alpha k_F}{T_K}\right)^2\nonumber\\
&+&\frac{31\zeta(5)}{12\pi^4}\left(\frac{2\alpha k_F}{T_K}\right)^4
-\dots\Big].
\end{eqnarray}
This yields the temperature $T_K$ as
\begin{eqnarray}
k_BT_K&=& \frac{2e^{\gamma}D}{\pi}\exp\Big[-\frac{1}{\rho_F|J|}
-\frac{7\zeta(3)}{2\pi^2}\left(\frac{2\alpha k_F}{k_BT_K}\right)^2\nonumber\\
&+&\frac{31\zeta(5)}{12\pi^4}\left(\frac{2\alpha k_F}{k_BT_K}\right)^4
-\cdots\Big],
\end{eqnarray}
where we introduced the Boltzmann constant $k_B$ and the density of states
$\rho_F=mk_F/(2\pi^2)$.
This is a self-consistency equation for $T_K$, and yields
\begin{eqnarray}
x&=& \exp\Big[ -0.21314\left(\frac{\alpha_r}{x}\right)^2
+0.0550\left(\frac{\alpha_r}{x}\right)^4\nonumber\\
&-&0.01655\left(\frac{\alpha_r}{x}\right)^6 
+0.005396\left(\frac{\alpha_r}{x}\right)^8 
-0.001822\left(\frac{\alpha_r}{x}\right)^{10}\nonumber\\
&+&\cdots \Big]\nonumber\\
&\equiv& g(x),
\end{eqnarray}
with variables
\begin{equation}
x= T_K/T_K^0,~~ \alpha_r=2\alpha k_F/k_BT_K^0,
\end{equation}
where
\begin{equation}
k_B T_K^0= \frac{2e^{\gamma}D}{\pi}\exp\left( -\frac{1}{\rho_F|J|}\right).
\end{equation}
We expanded $g(x)$ in powers of $\alpha_r/x$ up to the tenth order.
The function $g(x)$ is shown in Fig.1, where higher-order terms are small
and negligible except near $x\sim 0$.  
The equation $x=g(x)$ has no finite solution
when $\alpha_r>1.045$.  This indicates that
$T_K$ vanishes when the spin-orbit coupling $\alpha k_F$ is greater
than $1.045 k_BT_K^0$, and  
indicates a Kondo collapse with a sharp decrease of $T_K$.
This may overestimate the reduction of $T_K$.
When $\alpha$ is very large, if we use the asymptotic relation
$\psi(1/2+z)\sim \log(z)$, we obtain
\begin{equation}
1\simeq \rho_F|J|\Big[ \log\left(\frac{2e^{\gamma}D}{\pi k_BT}\right)
-\frac{1}{2}\log\left( \frac{2\alpha k_F}{\pi k_BT}\right)
+\frac{1}{4}\Big].
\end{equation}
This yields
\begin{equation}
k_B T_K\simeq \frac{\sqrt{e}}{2\alpha k_F}\frac{(2e^{\gamma}D)^2}{\pi}
\exp\left(-\frac{2}{\rho_F|J|}\right) 
= \frac{\pi\sqrt{e}}{\alpha_r}k_B T_K^0,
\end{equation}
for $\alpha_r\gg 1$.
We show $T_K$ as a function of $\alpha_r$ in Fig.2.

We expect that,in the strong coupling limit $\alpha_r\gg 1$, $T_K$
should approach that of single-band model:
\begin{equation}
k_B T_K^{\alpha} = \frac{2e^{\gamma}D}{\pi}\exp
\left(-\frac{2}{\rho_F|J|}\right).
\end{equation}
We will show this in the section 7.
This agrees with eq.(70) for $\alpha k_F\sim D$.
This is very small compared to the original $T_K$ because
$T_K^{\alpha}/T_K^0\simeq k_BT_K^0/D$.
Therefore $T_K$ decreases as $\alpha_r$ is increased and shows up
a sharp decrease at $\alpha_r\sim 1$.

\begin{figure}
\begin{center}
\includegraphics[width=\columnwidth]{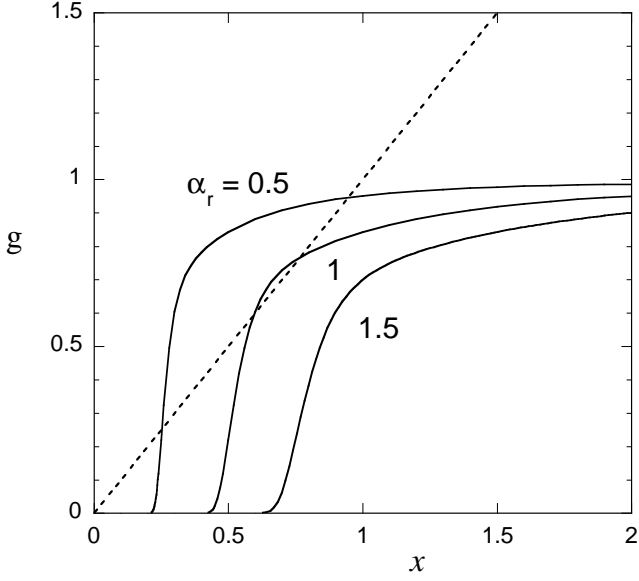}
\caption{
$g(x)=\exp(-(7\zeta(3)/2\pi^2)(\alpha_r/x)^2+\cdots)$ up to the tenth order
of $\alpha_r/x$ as a function of $x$
for $\alpha_r=0.5$, 1 and 1.5.  The straight line is a linear function $x$.
}
\end{center}
\label{g-x}
\end{figure}

\begin{figure}
\begin{center}
\includegraphics[width=\columnwidth]{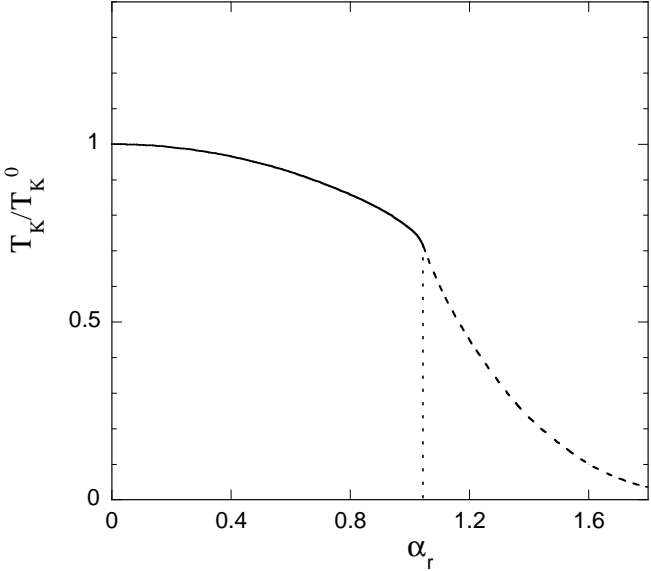}
\caption{
$T_K$ as a function of $\alpha_r$.
$x=g(x)$ has no solution for $\alpha_r>1.045$ (Kondo collapse) as
indicated by dotted line.
The dashed line is an expected line.
}
\end{center}
\label{tk-al}
\end{figure}

\section{Correction to Resistivity}

The imaginary part of $G_{{{\bf k}}{{\bf k}}}(\omega)^{-1}$ gives the scattering
rate of conduction electrons due to the localized spin.
The inverse of the life time $\tau_k(\omega)$ is
\begin{eqnarray}
\frac{1}{\tau_k(\omega)}&=& n_{i}N{\rm Im}G_{{{\bf k}}{{\bf k}}}(\omega)^{-1}
\nonumber\\
&=& \frac{3n_{i}J^2}{16}\pi\rho^{\alpha}(\omega)
\frac{1+JK(\omega)}{ (1+JK(\omega))^2+(JL(\omega))^2},
\nonumber\\
\end{eqnarray}
where $n_{i}$ is the concentration of magnetic impurities.
We have defined $K(\omega)={\rm Re}G(\omega+i\delta)$,
$L(\omega)=-{\rm Im}G(\omega+i\delta)$, and
$\rho^{\alpha}(\omega)=-(1/\pi){\rm Im}F(\omega+i\delta)$.
Because we obtain
\begin{equation}
K(0)\simeq \rho_F\Big[ \log\left(\frac{2e^{\gamma}D}{\pi k_BT}\right)
-\frac{7\zeta(3)}{4\pi^2}\left( \frac{2\alpha k_F}{k_BT}\right)^2\Big],
\end{equation}
the formula of the conductivity yields
\begin{eqnarray}
\sigma &=& -\frac{2e^2}{3}\int \tau_k(\xi_k)v_k^2
\frac{\partial f}{\partial \xi_k}\rho(\xi_k){\rm d}\xi_k\nonumber\\
&\simeq& \frac{2e^2}{3}v_F^2\rho(0)\frac{16}{3\pi n_{i}J^2\rho^{\alpha}(0)}
(1-|J|K(0))\nonumber\\
&\simeq& \frac{2e^2}{3}v_F^2\rho_F\frac{16}{3\pi n_i|J|}
\frac{\rho_F}{\rho^{\alpha}(0)}\Big[ \log\left(\frac{T}{T_K^0}\right)
\nonumber\\
&+& \frac{7\zeta(3)}{4\pi^2}\left(\frac{2\alpha k_F}{k_B T}\right)^2\Big].
\end{eqnarray}
We have the term $(\alpha/T)^2$  that comes from the spin-orbit 
interaction, and this term will conceal the logarithmic dependence
of the resistivity.
Then, the electrical resistivity $R$ in the high temperature region
$T\gg T_K^0$ is
\begin{eqnarray}
R &=& R_0\nonumber\\ 
&\times&\left[ 1-|J|\rho_F\log\left(\frac{2e^{\gamma}D}{\pi k_BT}
\right)+\frac{7\zeta(3)}{4\pi^2}|J|\rho_F\left(
\frac{2\alpha k_F}{k_BT}\right)^2 \right]^{-1}
\nonumber\\
\end{eqnarray}
where $R_0$ is a constant.
If the term $(\alpha/k_BT)^2$ is larger than the logarithmic term,
the resistivity even shows $R\sim T^2$.
Hence, the spin-orbit coupling may change the temperature dependence of the
resistivity drastically.

\section{Strong Spin-Orbit Coupling Case}

In this section let us consider the case with strong spin-orbit coupling.
For this purpose, we diagonalize the Hamiltonian $H_0$:
\begin{eqnarray}
H_0&=& \sum_{{\bf k}}(c_{{\bf k}\uparrow}^{\dag}c_{{\bf k}\downarrow}^{\dag})\left(
\begin{array}{cc}
\xi_{{\bf k}} & \alpha(ik_x+k_y) \\
\alpha(-ik_x+k_y) & \xi_{{\bf k}} \\
\end{array}
\right) \left(
\begin{array}{c}
c_{{\bf k}\uparrow} \\
c_{{\bf k}\downarrow} \\
\end{array}
\right)\nonumber\\
&=& \sum_{{\bf k}}\left[ (\xi_{{\bf k}}-\alpha k_{\perp})\alpha_{{\bf k}}^{\dag}
\alpha_{{\bf k}}+(\xi_{{\bf k}}+\alpha k_{\perp})\beta_{{\bf k}}^{\dag}
\beta_{{\bf k}} \right],
\end{eqnarray}
where $k_{\perp}=\sqrt{k_x^2+k_y^2}$, and $\alpha_{{\bf k}}$ and $\beta_{{\bf k}}$ are
defined by
\begin{eqnarray}
\alpha_{{\bf k}}&=&u_{{\bf k}}c_{{\bf k}\uparrow}+v_{{\bf k}}c_{{\bf k}\downarrow},
\\
\beta_{{\bf k}}&=&-v_{{\bf k}}^*c_{{\bf k}\uparrow}+u_{{\bf k}}c_{{\bf k}\downarrow}.
\end{eqnarray}
The coefficients $u_{{\bf k}}$ and $v_{{\bf k}}$ are
\begin{equation}
u_{{\bf k}}=\frac{1}{2},~~v_{{\bf k}}=-\frac{ik_x+k_y}{\sqrt{2}k_{\perp}},
\end{equation}
satisfying $u_{{\bf k}}^2+|v_{{\bf k}}|^2=1$.
We consider the case where the band
split is so large that we can neglect the upper band.
This means that we keep terms that contain $\alpha$-operators only.
In this approximation the s-d interaction term is
\begin{eqnarray}
H_{sd}^{\alpha}&=& -\frac{J}{2}\frac{1}{N}\sum_{{\bf k}{\bf k}'} [
S_z\{(u_ku_{k'}-v_kv_{k'}^*)\alpha_{{\bf k}}^{\dag}\alpha_{{{\bf k}}'} 
\nonumber\\
&+&S_+v_ku_{k'}\alpha_{{\bf k}}^{\dag}\alpha_{{{\bf k}}'}
+S_-u_kv_{k'}^*\alpha_{{\bf k}}^{\dag}\alpha_{{{\bf k}}'}].
\end{eqnarray}
This is the model of one-channel conduction-electron band that
interacts with the localized spin.

Let us consider the following Green's function:
\begin{equation}
G_{{\bf k}{\bf k}'}^{\alpha}(\tau)= -\langle T_{\tau}\alpha_{{\bf k}}(\tau)
\alpha_{{\bf k}'}^{\dag}(0)\rangle.
\end{equation}
By using the same method in previous sections, $G_{kk'}^{\alpha}$ is shown
to be
\begin{eqnarray}
G_{{{\bf k}}{{\bf k}}'}^{\alpha}(z)&=& \frac{\delta_{{{\bf k}}{{\bf k}}'}}
{z-\xi_{{{\bf k}}\alpha}}
+\frac{J}{2N}\frac{\frac{1}{2}+v_{{\bf k}}v_{{{\bf k}}'}^*}{(z-\xi_{{{\bf k}}\alpha})
(z-\xi_{{{\bf k}}'\alpha})}t(z),\nonumber\\
\end{eqnarray} 
for arbitrary complex number $z$ where we defined
\begin{equation}
t(z)= \frac{3J}{16}\frac{F_{\alpha}(z)}{1+\frac{J}{2}G_{\alpha}(z)
-\frac{3}{16}\left(\frac{J}{2}\right)^2F_{\alpha}(z)^2},
\end{equation}
\begin{eqnarray}
F_{\alpha}(z)&=& \frac{1}{N}\sum_{{\bf k}}\frac{1}{z-\xi_{{{\bf k}}\alpha}},\\
G_{\alpha}(z)&=& \frac{1}{N}\sum_{{\bf k}}\frac{\bar{n}_{{{\bf k}}\alpha}-1/2}
{z-\xi_{{{\bf k}}\alpha}}.
\end{eqnarray}
We derive this formula in Appendix.
The Kondo temperature $T_K^{\alpha}$ is determined from a singularity of $t(z)$
in the same way as previous sections.  We obtain 
\begin{equation}
k_B T_K^{\alpha}= \frac{2e^{\gamma}D}{\pi}
\exp\left( -\frac{2}{|J|\rho_F}\right).
\end{equation}
The characteristic energy $T_K^{\alpha}$ is reduced significantly
compared to the conventional Kondo temperature by factor 2 in the
exponential function.
This factor appears because the number of
channel of the conduction electrons in this case is just half of the normal
Kondo system.
The resistivity is also calculated as
\begin{equation}
R= R_0\Big[ 1+\frac{\rho_F|J|}{2}\log\left(\frac{2e^{\gamma}D}{\pi k_BT}
\right)+\cdots\Big],
\end{equation}
with a factor $1/2$.

\section{Discussion}

We investigated the Kondo effect in the presence of spin-orbit coupling.
The influence of band splitting was examined by using the Green's
function method where we adopted the decoupling scheme to obtain an
approximate solution.
The Kondo temperature is reduced by the spin-orbit interaction, and
shows a sudden decrease when $\alpha k_F$ is of the order of $k_BT_K^0$.
We call this the Kondo collapse due to the spin-orbit coupling.
The Kondo effect is suppressed
and the $\log T$-dependence of the resistivity will be weakened and concealed.
The reduction of $T_K$ as a result of the spin-orbit coupling is
consistent with the result for the single-impurity Anderson model using
the numerical renormalization group technique\cite{zit11}.
In their work the Kondo temperature is a decreasing function of the
Rashba energy $E_R\propto \alpha^2$ when the level of the localized 
electrons is lowered, that is, in the Kondo region, while it is a
increasing function when the localized level is not deep. 
The variation of the Kondo temperature is approximately linear as a
function of $E_R$, namely, quadratic in $\alpha$.
This is consistent with our result which shows a small variation of the
Kondo temperature with the quadratic correction when $\alpha$ is small.
In a recent work\cite{zar12}, the Kondo temperature is increased 
in the presence of the Dzyaloshinski-Moriya interaction.
The Dzyaloshinsky-Moriya interaction, however, vanishes in the
Kondo region with particle-hole symmetry $\epsilon_d=-U/2$.
Hence the result in ref.\cite{zar12} seems consistent with the result
for the s-d model.

As a limit of strong spin-orbit interaction, we can investigate a
crossover to a one-channel Kondo problem.
The Kondo problem with the spin-orbit coupling is closely
related to a multi-channel Kondo problem.
The Kondo temperature is reduced considerably because the degeneracy
of the conducting electrons
becomes half of the conventional Kondo system in this limit.
The specific heat also exhibits a $\log T$-term in the present model
with one-channel conduction band,
and this term appears in the fifth-order of $\rho J$. 
This agrees with the original Kondo problem.

The author expresses his sincere thanks to K. Yamaji, I. Hase and J. Kondo
for helpful discussion.

\section{Appendix}

In this appendix we derive the equation of motion for Green's functions
for the single-band s-d model and discuss its physical properties.

\subsection{Green's functions}

$H_0$ was diagonalized by $\alpha_{{\bf k}}$ and $\beta_{{\bf k}}$:
\begin{eqnarray}
\alpha_{{\bf k}}&=&u_{{\bf k}}c_{{\bf k}\uparrow}+v_{{\bf k}}c_{{\bf k}\downarrow},
\\
\beta_{{\bf k}}&=&-v_{{\bf k}}^*c_{{\bf k}\uparrow}+u_{{\bf k}}c_{{\bf k}\downarrow}.
\end{eqnarray}
The coefficients $u_{{\bf k}}$ and $v_{{\bf k}}$ are
\begin{equation}
u_{{\bf k}}=\frac{1}{2},~~v_{{\bf k}}=-\frac{ik_x+k_y}{\sqrt{2}k_{\perp}},
\end{equation}
satisfying $u_{{\bf k}}^2+|v_{{\bf k}}|^2=1$.
The s-d interaction part becomes 
\begin{eqnarray}
H_{sd}&=& -\frac{J}{2}\frac{1}{N}\sum_{{\bf k}{\bf k}'} [
S_z\{(u_ku_{k'}-v_kv_{k'}^*)\alpha_k^{\dag}\alpha_{k'}\nonumber\\
&-&(u_ku_{k'}-v_k^*v_{k'})\beta_{{{\bf k}}}^{\dag}\beta_{{{\bf k}}'} \nonumber\\
&-&(u_kv_{k'}+v_ku_{k'})\alpha_{{\bf k}}^{\dag}\beta_{{{\bf k}}'} 
- (u_kv_{k'}^*+v_k^*u_{k'})\beta_{{\bf k}}^{\dag}\alpha_{{{\bf k}}'}\}\nonumber\\
&+&S_+(v_ku_{k'}\alpha_{{\bf k}}^{\dag}\alpha_{{{\bf k}}'}-u_kv_{k'}
\beta_{{\bf k}}^{\dag}\beta_{{{\bf k}}'}
-v_kv_{k'}\alpha_{{\bf k}}^{\dag}\beta_{{{\bf k}}'}\nonumber\\
&+&u_ku_{k'}\beta_{{\bf k}}^{\dag}\alpha_{{{\bf k}}'})
\nonumber\\
&+&S_-(u_kv_{k'}^*\alpha_{{\bf k}}^{\dag}\alpha_{{{\bf k}}'}-v_k^*u_{k'}
\beta_{{\bf k}}^{\dag}\beta_{{{\bf k}}'}
+u_ku_{k'}\alpha_{{\bf k}}^{\dag}\beta_{{{\bf k}}'}\nonumber\\
&-&v_k^*v_{k'}\beta_{{\bf k}}^{\dag}\alpha_{{{\bf k}}'})].
\end{eqnarray}
The single-band s-d model contains only the following s-d interaction,
\begin{eqnarray}
H_{sd}^{\alpha}&=& -\frac{J}{2}\frac{1}{N}\sum_{{\bf k}{\bf k}'} [
S_z\{(u_ku_{k'}-v_kv_{k'}^*)\alpha_k^{\dag}\alpha_{k'} \nonumber\\
&+&S_+v_ku_{k'}\alpha_k^{\dag}\alpha_{k'}
+S_-u_kv_{k'}^*\alpha_k^{\dag}\alpha_{k'}].
\end{eqnarray}

We consider the following Green's functions:
\begin{eqnarray}
G_{{\bf k}{\bf k}'}^{\alpha}(\tau)&=& -\langle T_{\tau}\alpha_{{\bf k}}(\tau)
\alpha_{{\bf k}'}^{\dag}(0)\rangle,\\
\langle\langle S_z\alpha_{{\bf k}};\alpha_{{\bf k}'}^{\dag}\rangle\rangle_{\tau}
&=& -\langle T_{\tau}(S_z\alpha_{{\bf k}})(\tau)\alpha_{{\bf k}'}^{\dag}(0)
\rangle,\\
\langle\langle S_+\alpha_{{\bf k}};\alpha_{{\bf k}'}^{\dag}\rangle\rangle_{\tau}
&=& -\langle T_{\tau}(S_+\alpha_{{\bf k}})(\tau)\alpha_{{\bf k}'}^{\dag}(0)
\rangle.
\end{eqnarray}
The Fourier transforms are defined similarly:
\begin{eqnarray}
G_{{\bf k}{\bf k}'}^{\alpha}(\tau)&=& \frac{1}{\beta}\sum_{\omega_n}
e^{i\omega_n\tau}G_{{\bf k}{\bf k}'}^{\alpha}(i\omega_n),\\
\langle\langle S_z\alpha_{{\bf k}};\alpha_{{\bf k}'}^{\dag}\rangle\rangle_{\tau}
&=& \frac{1}{\beta}\sum_{\omega_n}e^{i\omega_n\tau}
\langle\langle S_z\alpha_{{\bf k}};\alpha_{{\bf k}'}^{\dag}\rangle\rangle_{\omega},
\\
\langle\langle S_+\alpha_{{\bf k}};\alpha_{{\bf k}'}^{\dag}\rangle\rangle_{\tau}
&=& \frac{1}{\beta}\sum_{\omega_n}e^{i\omega_n\tau}
\langle\langle S_+\alpha_{{\bf k}};\alpha_{{\bf k}'}^{\dag}\rangle\rangle_{\omega}.
\end{eqnarray}
The equations of motion for these Green's functions are derived as follows.
\begin{eqnarray}
i\omega_n G_{{{\bf k}}{{\bf k}}'}^{\alpha}(i\omega_n)&=& \delta_{{{\bf k}}{{\bf k}}'}
+\xi_{{{\bf k}}\alpha}
G_{{{\bf k}}{{\bf k}}'}^{\alpha}(i\omega_n)\nonumber\\
&-&\frac{J}{2N}\sum_q[ (u_ku_q-v_kv_q^*)\langle\langle S_z\alpha_{{\bf q}};
\alpha_{{{\bf k}}'}^{\dag}\rangle\rangle_{i\omega_n} \nonumber\\
&+& v_ku_q\langle\langle S_+\alpha_{{\bf q}};\alpha_{{{\bf k}}'}^{\dag}
\rangle\rangle_{i\omega_n}\nonumber\\
&+& u_kv_q^*\langle\langle S_-\alpha_{{\bf q}};\alpha_{{{\bf k}}'}^{\dag}
\rangle\rangle_{i\omega_n}
],
\end{eqnarray}

\begin{eqnarray}
&&i\omega_n\langle\langle S_z\alpha_{{\bf k}};\alpha_{{{\bf k}}'}^{\dag}
\rangle\rangle_{i\omega_n}
= \xi_{{{\bf k}}\alpha}\langle\langle S_z\alpha_{{\bf k}};\alpha_{{{\bf k}}'}^{\dag}
\rangle\rangle_{i\omega_n}\nonumber\\
&+&\frac{J}{2N}\sum_q[ -(u_ku_q-v_kv_q^*)
\langle\langle S_z^2\alpha_{{\bf q}};\alpha_{{{\bf k}}'}^{\dag}
\rangle\rangle_{i\omega_n}
\nonumber\\
&+& v_ku_q^*
\langle\langle S_+(n_{{{\bf k}}\alpha}-\frac{1}{2})\alpha_{{\bf q}};
\alpha_{{{\bf k}}'}^{\dag}
\rangle\rangle_{i\omega_n} \nonumber\\
&-& u_kv_q^*
\langle\langle S_-(n_{{{\bf k}}\alpha}-\frac{1}{2})\alpha_{{\bf q}};
\alpha_{{{\bf k}}'}^{\dag}
\rangle\rangle_{i\omega_n} ],\nonumber\\
\end{eqnarray}

\begin{eqnarray}
&&i\omega_n\langle\langle S_+\alpha_{{\bf k}};\alpha_{{{\bf k}}'}^{\dag}
\rangle\rangle_{i\omega_n}
= \xi_{{{\bf k}}\alpha}\langle\langle S_+\alpha_{{\bf k}};\alpha_{{{\bf k}}'}^{\dag}
\rangle\rangle_{i\omega_n}\nonumber\\
&+&\frac{J}{2N}\sum_{{\bf q}}[ -(u_ku_q-v_kv_q^*)
\langle\langle S_+(n_{{{\bf k}}\alpha}-\frac{1}{2})\alpha_{{\bf q}};
\alpha_{{{\bf k}}'}^{\dag}
\rangle\rangle_{i\omega_n}
\nonumber\\
&+& 2u_kv_q^*
\langle\langle S_zn_{{{\bf k}}\alpha}\alpha_{{\bf q}};\alpha_{{{\bf k}}'}^{\dag}
\rangle\rangle_{i\omega_n} 
- u_kv_q^*
\langle\langle S_+S_-\alpha_{{\bf q}};\alpha_{{{\bf k}}'}^{\dag}
\rangle\rangle_{i\omega_n} ],
\nonumber\\
\end{eqnarray}

\begin{eqnarray}
&&i\omega_n\langle\langle S_-\alpha_{{\bf k}};\alpha_{{{\bf k}}'}^{\dag}
\rangle\rangle_{i\omega_n}
= \xi_{{{\bf k}}\alpha}\langle\langle S_+\alpha_{{\bf k}};\alpha_{{{\bf k}}'}^{\dag}
\rangle\rangle_{i\omega_n}\nonumber\\
&+&\frac{J}{2N}\sum_{{\bf q}}[ (u_ku_q-v_kv_q^*)
\langle\langle S_-(n_{{{\bf k}}\alpha}-\frac{1}{2})\alpha_{{\bf q}};
\alpha_{{{\bf k}}'}^{\dag}
\rangle\rangle_{i\omega_n}
\nonumber\\
&-& 2v_ku_q
\langle\langle S_zn_{{{\bf k}}\alpha}\alpha_{{\bf q}};\alpha_{{{\bf k}}'}^{\dag}
\rangle\rangle_{i\omega_n} 
- v_ku_q
\langle\langle S_-S_+\alpha_{{\bf q}};\alpha_{{{\bf k}}'}^{\dag}
\rangle\rangle_{i\omega_n} ],
\nonumber\\
\end{eqnarray}
where $n_{{{\bf k}}\alpha}=\alpha_{{\bf k}}^{\dag}\alpha_{{\bf k}}$.
We have unknown functions
$\langle\langle S_a n_{{{\bf k}}\alpha}\alpha_{{\bf q}};\alpha_{{{\bf k}}'}^{\dag}
\rangle\rangle$
for $a=z$, $+$ and $-$.

\subsection{Approximate Solution}

To obtain a consistent solution, we adopt the following approximation:
\begin{equation}
\langle\langle S_a n_{{{\bf k}}\alpha}\alpha_{{\bf q}};\alpha_{{{\bf k}}'}^{\dag}
\rangle\rangle
=\langle n_{{{\bf k}}\alpha}\rangle\langle\langle S_a\alpha_{{\bf q}};
\alpha_{{{\bf k}}'}^{\dag}
\rangle\rangle.
\end{equation}
Using this approximation and the relation $S_+S_-=3/4+S_z-S_z^2$, 
we obtain
\begin{eqnarray}
&&(i\omega_n-\xi_{{{\bf k}}\alpha})G_{{{\bf k}}{{\bf k}}'}^{\alpha}(i\omega_n)= 
\delta_{{{\bf k}}{{\bf k}}'}\nonumber\\
&+& \left(\frac{J}{2N}\right)^2\sum_{{\bf q}}\frac{\bar{n}_{{\bf q}}-1/2}
{i\omega_n-\xi_{{\bf q}}}
\sum_{{{\bf q}}'}\Big[ v_ku_{q'}\langle\langle S_+\alpha_{{{\bf q}}'};
\alpha_{{{\bf k}}'}^{\dag}
\rangle\rangle_{i\omega_n}\nonumber\\
&+& u_kv_{q'}^*\langle\langle S_-\alpha_{{{\bf q}}'};\alpha_{{{\bf k}}'}^{\dag}
\rangle\rangle_{i\omega_n}\nonumber\\
&+&(u_ku_{q'}-v_kv_{q'}^*)\langle\langle S_z\alpha_{{{\bf q}}'};
\alpha_{{{\bf k}}'}^{\dag}
\rangle\rangle_{i\omega_n}\Big]\nonumber\\
&+&\frac{3}{8}\left(\frac{J}{2N}\right)^2
\sum_{{\bf q}}\frac{1}{i\omega_n-\xi_{{{\bf q}}\alpha}}\sum_{{{\bf q}}'}
(u_ku_{q'}+v_kv_{q'}^*)
G_{{{\bf q}}'{{\bf k}}'}^{\alpha}(i\omega_n),\nonumber\\
\end{eqnarray}
where $\bar{n}_{{\bf q}}=\langle n_{{{\bf q}}\alpha}\rangle$.
Here we define
\begin{eqnarray}
\Gamma_{{{\bf k}}{{\bf k}}'}(i\omega_n)&=&\sum_{{\bf q}} \Big[(u_ku_q-v_kv_q^*)
\langle\langle S_z\alpha_{{\bf q}};
\alpha_{{{\bf k}}'}^{\dag}\rangle\rangle_{i\omega_n}\nonumber\\
&+&v_ku_q\langle\langle S_+\alpha_{{\bf q}};\alpha_{{{\bf k}}'}^{\dag}
\rangle\rangle_{i\omega_n}+u_kv_q^*\langle\langle S_-\alpha_{{\bf q}};
\alpha_{{{\bf k}}'}^{\dag}\rangle\rangle_{i\omega_n}\Big].\nonumber\\
\end{eqnarray}
This quantity reads after substituting the equations for
$\langle\langle S_a\alpha_{{\bf q}};\alpha_{{{\bf k}}'}^{\dag}\rangle\rangle$
\begin{eqnarray}
\Gamma_{{{\bf k}}{{\bf k}}'}&=& \frac{J}{2N}\sum_{{\bf q}}\frac{1}
{i\omega_n-xi_{{{\bf k}}\alpha}}
\sum_{{{\bf q}}'}\Big[\nonumber\\
&-&v_ku_{q'}\left(\bar{n}_{{\bf q}}-\frac{1}{2}\right)
\langle\langle S_+\alpha_{{{\bf q}}'};\alpha_{{{\bf k}}'}^{\dag}
\rangle\rangle_{i\omega_n}
\nonumber\\
&-& u_kv_{q'}^*\left(\bar{n}_{{\bf q}}-\frac{1}{2}\right)
\langle\langle S_-\alpha_{{{\bf q}}'};\alpha_{{{\bf k}}'}^{\dag}
\rangle\rangle_{i\omega_n}\nonumber\\
&-&(u_ku_{q'}-v_kv_{q'}^*)\left(\bar{n}_{{\bf q}}-\frac{1}{2}\right)
\langle\langle S_z\alpha_{{{\bf q}}'};\alpha_{{{\bf k}}'}^{\dag}
\rangle\rangle_{i\omega_n}
\nonumber\\
&-&\frac{3}{8}(u_ku_{q'}+v_kv_{q'}^*)G_{{{\bf q}}'{{\bf k}}'}^{\alpha}(i\omega_n) 
\Big]\nonumber\\
&=& -\frac{J}{2N}\frac{3}{8}\sum_{{\bf q}}\frac{\bar{n}_{{\bf k}}-1/2}
{i\omega_n-\xi_{{{\bf q}}\alpha}}
\Gamma_{{{\bf k}}{{\bf k}}'}
-\frac{J}{2N}\frac{3}{8}\sum_{{\bf q}}\frac{1}{i\omega_n-\xi_{{{\bf q}}\alpha}}
\nonumber\\
&\times&\sum_{{{\bf q}}'}(u_ku_{q'}+v_kv_{q'}^*)
G_{{{\bf q}}'{{\bf k}}'}^{\alpha}(i\omega_n).
\end{eqnarray}
Then we obtain
\begin{eqnarray}
G_{{{\bf k}}{{\bf k}}'}^{\alpha}(i\omega_n)&=& \frac{\delta_{{{\bf k}}{{\bf k}}'}}
{i\omega_n-\xi_{{{\bf k}}\alpha}}
+\frac{3}{8}\left(\frac{J}{2N}\right)^2\frac{1}{i\omega_n-\xi_{{{\bf k}}\alpha}}
\nonumber\\
&\times&\sum_{{{\bf q}}'}\frac{1}{i\omega_n-\xi_{{{\bf q}}'\alpha}}
\frac{1}{1+\frac{J}{2N}\sum_{{{\bf p}}}
\frac{\bar{n}_{{\bf p}}-1/2}{i\omega_n-\xi_{{{\bf p}}\alpha}}}\nonumber\\
&\times& \sum_{{\bf q}} (u_ku_q+v_kv_q^*)G_{{{\bf q}}{{\bf k}}'}^{\alpha}(i\omega_n).
\end{eqnarray}
We have set $u_{{\bf k}}=1/\sqrt{2}$.  Because $v_{{\bf k}}$ satisfies 
$v_{{\bf k}}=-v_{-{\bf k}}$ and $|v_{{\bf k}}|^2=1/2$,  we have
\begin{eqnarray}
\sum_{{\bf k}}v_{{\bf k}}^*G_{{{\bf k}}{{\bf k}}'}^{\alpha}(i\omega_n)&=&
\frac{v_{{\bf k}'}^*}{i\omega_n-\xi_{{{\bf k}}'\alpha}}\Big[1 
-\frac{3}{8}
\left(\frac{J}{2}\right)^2\frac{1}{2}F_{\alpha}(i\omega_n)^2\nonumber\\
&\times&\frac{1}{1+(J/2)G_{\alpha}(i\omega_n)} \Big]^{-1},
\end{eqnarray}
where we set
\begin{eqnarray}
F_{\alpha}(z)&=& \frac{1}{N}\sum_{{\bf k}}\frac{1}{z-\xi_{{{\bf k}}\alpha}},\\
G_{\alpha}(z)&=& \frac{1}{N}\sum_{{\bf k}}\frac{\bar{n}_{{{\bf k}}\alpha}-1/2}
{z-\xi_{{{\bf k}}\alpha}}.
\end{eqnarray}
We define
\begin{equation}
t(z)= \frac{3J}{16}\frac{F_{\alpha}(z)}{1+\frac{J}{2}G_{\alpha}(z)
-\frac{3}{16}\left(\frac{J}{2}\right)^2F_{\alpha}(z)^2}.
\end{equation}
Then $G_{{{\bf k}}{{\bf k}}'}^{\alpha}$ and $\Gamma_{{{\bf k}}{{\bf k}}'}$ read
\begin{eqnarray}
G_{{{\bf k}}{{\bf k}}'}^{\alpha}(z)&=& \frac{\delta_{{{\bf k}}{{\bf k}}'}}
{z-\xi_{{{\bf k}}\alpha}}
+\frac{J}{2N}\frac{\frac{1}{2}+v_kv_{k'}^*}{(z-\xi_{{{\bf k}}\alpha})
(z-\xi_{{{\bf k}}'\alpha})}t(z),\nonumber\\
&&\\
\Gamma_{{{\bf k}}{{\bf k}}'}(z)&=& -\frac{\frac{1}{2}+v_kv_{k'}^*}
{z-\xi_{{{\bf k}}'\alpha}}t(z),
\end{eqnarray}
for arbitrary complex number $z$.
The Kondo temperature $T_K^{\alpha}$ is determined from a singularity of $t(z)$
in the same way as previous sections.  We obtain
\begin{equation}
T_K^{\alpha}= \frac{2e^{\gamma}D}{\pi}
\exp\left( -\frac{2}{|J|\rho_F}\right).
\end{equation}
The characteristic energy $T_K^{\alpha}$ is reduced significantly
compared to the conventional Kondo temperature by factor 2 in the
exponential function:  
\begin{equation}
T_K^{\alpha}\sim \left(\frac{T_K^0}{D}\right)T_K^0.
\end{equation}
This factor appears because the number of 
channel of the conduction electrons in this case is just half of the normal
Kondo system. 
The resistivity is also calculated as
\begin{equation}
R= R_0\Big[ 1+\frac{\rho_F|J|}{2}\log\left(\frac{2e^{\gamma}D}{\pi k_BT}
\right)+\cdots\Big],
\end{equation}
with a factor $1/2$.

\subsection{Entropy and Specific Heat}

The energy expectation value $E=\langle H\rangle$ is given by
\begin{eqnarray}
E&=& \sum_{{\bf k}}\xi_{{\bf k}}\langle\alpha_{{\bf k}}^{\dag}\alpha_{{\bf k}}\rangle
-\frac{J}{2N}
\sum_{{{\bf k}}{{\bf k}}'}\langle\{S_z(u_ku_{k'}-v_kv_{k'}^*)\nonumber\\
&+& S_+v_ku_{k'}
+S_-u_kv_{k'}^*\}\alpha_{{\bf k}}^{\dag}\alpha_{{{\bf k}}'}\rangle\nonumber\\
&=& \frac{1}{\beta}\sum_{{{\bf k}}\omega_n}\xi_{{{\bf k}}\alpha}
G_{{{\bf k}}{{\bf k}}}^{\dag}(i\omega_n)
-\frac{J}{2}\frac{1}{\beta N}\sum_{{\bf k}}\Gamma_{{{\bf k}}{{\bf k}}}(i\omega_n)
\nonumber\\
&=& \frac{1}{\beta}\sum_{{{\bf k}}\omega_n}
\frac{\xi_{{{\bf k}}\alpha}}{i\omega_n-\xi_{{{\bf k}}\alpha}}
+\frac{J}{2}\frac{1}{\beta N}\sum_{{{\bf k}}\omega_n}\frac{i\omega_n t(i\omega_n)}
{(i\omega_n-\xi_{{{\bf k}}\alpha})^2}.\nonumber\\
\end{eqnarray}
The expectation value of the interaction Hamiltonian is denoted as $V$.
$V$ is given by
\begin{eqnarray}
V&=& -\frac{J}{2N}
\sum_{{{\bf k}}{{\bf k}}'}\langle\{S_z(u_ku_{k'}-v_kv_{k'}^*)
+ S_+v_ku_{k'}\nonumber\\
&+& S_-u_kv_{k'}^*\}\alpha_{{\bf k}}^{\dag}\alpha_{{{\bf k}}'}\rangle\nonumber\\
&=&
-\frac{J}{2}\frac{1}{\beta N}\sum_{{\bf k}}\Gamma_{{{\bf k}}{{\bf k}}}(i\omega_n)
=
\frac{J}{2}\frac{1}{\beta N}\sum_{{{\bf k}}\omega_n}\frac{t(i\omega_n)}
{i\omega_n-\xi_{{{\bf k}}\alpha}}.\nonumber\\
\end{eqnarray}
This is written as
\begin{equation}
V= \frac{J}{2}\rho(0){\rm Re}\int_{-D}^D d\omega f(\omega)
t(\omega-i\delta),
\end{equation}
where we adopted the approximation
\begin{equation}
F_{\alpha}(\omega\pm i\delta)= \mp\pi\rho(0)i.
\end{equation}
$\rho(\omega)$ is the density of states of conduction electrons.

We need $t(z)$ to estimate $V$.
$G_{\alpha}(z)$, which appears in the denominator of $t(z)$, contains
a singularity.  $G_{\alpha}(z)$ is written as
\begin{eqnarray}
G_{\alpha}(z)&=& R_{\alpha}(z)+\frac{J}{2N}\frac{1}{\beta N}\sum_{{{\bf k}}\omega_n}
\frac{1}{z-\xi_{{{\bf k}}\alpha}}\frac{t(i\omega_n)}{(i\omega_n-
\xi_{{{\bf k}}\alpha})^2},
\nonumber\\
\end{eqnarray}
where
\begin{equation}
R_{\alpha}(z)= \frac{1}{\beta}\sum_{\omega_n}
\frac{F_{\alpha}(i\omega_n)-F_{\alpha}(z)}{z-i\omega_n}-\frac{1}{2}
F_{\alpha}(z).
\end{equation}
$R_{\alpha}(z)$ is evaluated as\cite{zit68}
\begin{equation}
R_{\alpha}(\omega-i\delta)\approx \rho(0)\left[ \psi\left(\frac{1}{2}+
\frac{\beta D}{2\pi}\right)-\psi\left(\frac{1}{2}+i\frac{\beta z}{2\pi}
\right)\right],
\end{equation}
where $\psi$ is the digamma function and $D$ is the cutoff energy.  
We use the following relation,
\begin{eqnarray}
&&1+\frac{\rho(0)J}{2}\Big[ \log\left(\frac{D}{2\pi k_BT}\right)
-\psi\left(\frac{1}{2}+i\frac{\beta\omega}{2\pi}\right)\Big]\nonumber\\
&=& \frac{\rho(0)J}{2}\Big[ \log\frac{T_K^{\alpha}}{T}-g(\beta\omega)\Big],
\end{eqnarray}
where
\begin{equation}
g(x)= \psi\left( \frac{1}{2}+i\frac{x}{2\pi}\right)-\psi\left(\frac{1}{2}
\right).
\end{equation}
Then the interaction energy is
\begin{equation}
V= -\frac{3\pi}{16}\rho(0)J{\rm Im}\int_{-D}^D d\omega
f(\omega)\frac{1}
{\log(T_K^{\alpha}/T)-g(\beta\omega)},
\end{equation}
where we neglected the term of the order of $(\rho(0)J)^2$ in the
denominator of $t(z)$.
$V$ has a logarithmic temperature dependence.
Because of the relation between the free energy and $V$,
\begin{equation}
V = J\frac{\partial F}{\partial J},
\end{equation}
the additional entropy $\Delta S(T)$ is
\begin{equation}
\Delta S(T)= -\frac{\partial}{\partial T}(F-F_0)
=-\int_0^J\frac{dJ'}{J'}\frac{\partial}{\partial T}V(J',T).
\end{equation}
To estimate $V$, we use the expansion formula for the Fermi
distribution function $f(\omega)$:
\begin{equation}
\int_{-D}^D d\omega f(\omega)h(\omega)=\int_{-D}^0 d\omega h(\omega)
+\frac{\pi^2}{6}(k_BT)^2 h'(0),
\end{equation}
for a differentiable function $h(\omega)$.  Using this, we obtain
\begin{eqnarray}
V&=&-\frac{3\pi}{16}\rho(0)J {\rm Im}\int_{-D}^0 d\omega \frac{1}
{\log(T_K^{\alpha}/T)-g(\beta\omega)}
\nonumber\\
&-& \frac{3\pi}{16}\frac{\pi^2}{6}(k_BT)^2\rho(0)J{\rm Im} 
\frac{\partial}{\partial\omega}
\frac{1}{\log(T_K^{\alpha}/T)-g(\beta\omega)}\Big|_{\omega=0}.
\nonumber\\
\end{eqnarray}
We are interested in logarithmic terms $\log(D/k_BT)$, $\log(D/k_BT)^2$
and so on in the region $|\log(T_K^{\alpha}/T)|\gg 1$.
The second term is written as
\begin{equation}
V_2 = -\frac{\pi^4}{128}k_BT\rho(0)J\frac{1}{(\log(T_K^{\alpha}/T))^2}.
\end{equation}
This is expanded as in terms of $\rho(0)J$:
\begin{eqnarray}
V_2&=& \frac{\pi^4}{32}\left(\frac{\rho(0)J}{2}\right)^4 k_BT
\log\left(\frac{D}{k_BT}\right)\nonumber\\
&-& \frac{3\pi^4}{64}\left(\frac{\rho(0)J}{2}\right)^5 k_BT
\log\left(\frac{D}{k_BT}\right)^2.
\end{eqnarray}
In the first term of $V$, the logarithmic corrections never emerge
from the region where $\beta\omega$ is large
because we have $g(\beta\omega)\sim \log(\beta\omega)$ for large
$\omega$ and the $T$-dependence is canceled with $\log(T_K^{\alpha}/T)$.
When $\beta\omega$ is small, $g(\beta\omega)$ is expressed in a
power series of $\beta\omega$.  A dominant contribution is of the
order of $(\log(T_K^{\alpha}/T))^{-2}$.  The integral is restricted on
the interval $(-k_BT,0)$ and the first term $V_1$ is estimated as
\begin{eqnarray}
V_1&\simeq& -\frac{3\pi}{16}\rho(0)J\frac{1}{(\log(T_K^{\alpha}/T))^2}
\int_{-k_BT}^0d\omega {\rm Im}\psi\left(\frac{1}{2}+i\frac{\beta\omega}{2\pi}
\right)\nonumber\\
&=& -\frac{3\pi}{16}k_BT\rho(0)J\frac{1}{(\log(T_K^{\alpha}/T))^2}\pi
\Big[ -\frac{1}{8}+0.0052\nonumber\\
&-&0.00738+0.000026\cdots \Big].
\end{eqnarray}
As a result, $V$ is given as
\begin{equation}
V = -\frac{A}{2}k_BT\rho(0)J\frac{1}{(\log(T_K^{\alpha}/T))^2},
\end{equation}
for a constant $A>0$.

From the relation $T_K^{\alpha}=D\exp(2/(\rho(0)J))$, we have
\begin{equation}
\frac{d\rho(0)J}{\rho(0)J}=-\frac{1}{\log(T_K^{\alpha}/D)}
d\log T_K^{\alpha}.
\end{equation}
Using this formula, the entropy obtained from the interaction energy $V$ is
\begin{equation}
\Delta S= -\frac{\partial\Delta F}{\partial T},
\end{equation}
where
\begin{eqnarray}
\Delta F&=& -k_BA\Big[ T
\frac{1}{(\log(D/T))^2}\Bigl( \frac{\rho(0)J}{2}
+ \frac{1}{\log(T_K^{\alpha}/T)}\nonumber\\
&-& \frac{2}{\log(D/T)}
\log\Big|\frac{\rho(0)J}{2}\log\left(\frac{T_K^{\alpha}}{T}\Bigr)
\Big|\right)\Big],
\end{eqnarray}
is the free energy.
Because of the relation
\begin{equation}
\log\left(\frac{T_K^{\alpha}}{T}\right)= \frac{2}{\rho(0)J}
+\log\left(\frac{2e^{\gamma}}{\pi}\frac{D}{k_BT}\right),
\end{equation}
up to the fifth order of $\rho(0)J$, $\Delta S$ is given as
\begin{eqnarray}
\Delta S &=& k_B A\Big[ \frac{1}{3}
\left(\frac{\rho(0)J}{2}\right)^3
+\frac{1}{2}\left(\frac{\rho(0)J}{2}\right)^4\nonumber\\
&-& \frac{1}{2}\left(\frac{\rho(0)J}{2}\right)^4
\log\left( \frac{D}{k_BT}\right)
+ \frac{3}{5}\left(\frac{\rho(0)J}{2}\right)^5
\left(\log\frac{D}{k_BT}\right)^2\nonumber\\
&-& \frac{6}{5}\left(\frac{\rho(0)J}{2}\right)^5
\log\left( \frac{D}{k_BT}\right)\Big].
\end{eqnarray}
The logarithmic term first appears in the fourth order of $\rho(0)J$.
Then the correction to the specific heat 
$\Delta C=T\partial \Delta S/\partial T$ is
\begin{equation}
\frac{\Delta C}{k_B}\simeq \frac{A}{2}
\left(\frac{\rho(0)J}{2}\right)^4\Big[ 1
-\frac{12}{5}\left(\frac{\rho(0)J}{2}\right)
\log\left(\frac{D}{k_BT}\right)\Big].
\end{equation}
Hence the specific heat exhibits a logarithmic behavior at low temperatures.
A $\log T$-term appears in the fifth order of $\rho(0)J$; this agrees
with the original Kondo problem\cite{kon69}.
In the original Kondo problem, the entropy and the specific heat were evaluated
as\cite{kon69,yos69}
\begin{eqnarray}
\Delta S_{sd}&\simeq& k_B\frac{\pi^2}{4}(\rho J)^3\Big[1-3\rho J\log\left(
\frac{D}{k_B T}\right)\Big],\\
\Delta C_{sd}&\simeq& k_B\frac{3\pi^3}{4}(\rho J)^4\Big[1
-4\rho J\log\left( \frac{D}{k_B T}\right)\Big].
\end{eqnarray}
This suggests that\cite{zit68}
\begin{eqnarray}
\Delta C_{sd}&\simeq& k_B\frac{3\pi^3}{4}(\rho J)^4
\frac{1}{(1+\rho J\log(D/k_BT)^4}\nonumber\\
&\simeq& k_B\frac{3\pi^3}{4}\frac{1}{(\log(T_K/T))^4},
\end{eqnarray}
as an expansion in terms of $1/\log(T_K/T)$.
In the present model, the coefficients are reduced, where 4 is reduced to 12/5
in front of $\rho J\log(D/k_BT)$ in the specific heat 
compared to the usual s-d model,
and the divergence near the Kondo temperature is moderated.
Because the formation of a local singlet by the
the conduction electrons is weakened in a one-channel case, the 
entropy decreases more slowly as the temperature is decreased.

In the region $|\log(D/k_BT)|\gg 1$ and $|\log(T_K^{\alpha}/T)|\gg 1$,
$\Delta S$ is obtained as a double-power series of $1/\log(D/k_BT)$
and $1/\log(T_K^{\alpha}/T)$:
\begin{equation}
\Delta S\simeq k_BA\Big[ \frac{1}{(\log(D/k_BT))^2}\left(
\frac{\rho(0)J}{2}+\frac{1}{\log(T_K^{\alpha}/T)}\right)\Big].
\end{equation}
Then we obtain
\begin{equation}
\Delta C\simeq k_BA\frac{1}{(\log(D/k_BT))^2}\frac{1}{(\log(T_K^{\alpha}/T))^2}.
\end{equation}

\end{document}